\documentclass[journal]{IEEEtran}

\makeatletter
\def\bstctlcite{\@ifnextchar[{\@bstctlcite}{\@bstctlcite[@auxout]}}
\def\@bstctlcite[#1]#2{\@bsphack
  \@for\@citeb:=#2\do{%
    \edef\@citeb{\expandafter\@firstofone\@citeb}%
    \if@filesw\immediate\write\csname #1\endcsname{\string\citation{\@citeb}}\fi}%
  \@esphack}
\makeatother

\usepackage{inputenc}
\usepackage[cmex10]{amsmath}
\usepackage{verbatim}
\usepackage{array}
\usepackage{multicol}
\usepackage{multirow}
\usepackage{dcolumn}
\usepackage{color}
\usepackage[noadjust]{cite}
\usepackage{url}
\usepackage{balance}
\usepackage[usenames,dvipsnames]{xcolor}
\usepackage[flushleft]{threeparttable}
\usepackage{siunitx}
\usepackage{xspace}
\usepackage{microtype}
\usepackage{graphicx}
\usepackage{booktabs} 
\usepackage{cancel}

\usepackage{hyperref}
\usepackage{nicefrac}       
\usepackage{amssymb,amsfonts}       
\usepackage{stackengine,mathtools}
\usepackage{amsthm}

\usepackage{cleveref}
\crefformat{section}{\S#2#1#3}
\crefformat{subsection}{\S#2#1#3}
\crefformat{subsubsection}{\S#2#1#3}

\usepackage{xr}

\newcommand{\acro}[1]{\textsc{#1}\xspace}

\newcommand{\acd}{\acro{acd}}
\newcommand{\ncsc}{\acro{ncsc}}
\newcommand{\dstl}{\acro{dstl}}
\newcommand{\ai}{\acro{ai}}
\newcommand{\ml}{\acro{ml}}
\newcommand{\rl}{\acro{rl}}
\newcommand{\cgm}{\acro{cgm}}
\newcommand{\mitre}{\acro{MITRE ATT\&CK}}

\begin{document}
\bstctlcite{IEEEexample:BSTcontrol}
\title{Prospective Artificial Intelligence Approaches \\ for Active Cyber Defence}

\author{Neil Dhir$^{a,*}$, Henrique Hoeltgebaum$^{b,*}$, Niall Adams$^{b}$, Mark Briers$^{a,b}$, Anthony Burke$^{a}$ and Paul Jones$^{a}$
\thanks{The authors would like to acknowledge funding from the UK Government, through the Defence \& Security partnership at The Alan Turing Institute. Also the authors would like to acknowledge inputs from Chris W from UK's Defence Science and Technology Laboratory respectively. Also we are grateful to four anonymous reviewers for their insightful comments. Corresponding authors: \texttt{ndhir@turing.ac.uk} and \texttt{hh3015@imperial.ac.uk}.}
\thanks{
$^{a}$ The Alan Turing Institute, UK \\
$^{b}$ Imperial College London, Departament of Mathematics, UK \\
$^*$Equal contribution.}
}

\maketitle
\begin{abstract}
	Cyber criminals are rapidly developing new malicious tools that leverage artificial intelligence (AI) to enable new classes of adaptive and stealthy attacks. New defensive methods need to be developed to counter these threats. Some cyber security professionals are speculating AI will enable corresponding new classes of active cyber defence measures -- is this realistic, or currently mostly hype? The Alan Turing Institute,  with expert  guidance from the UK National Cyber Security Centre  and Defence Science Technology Laboratory, published a research roadmap for AI for ACD last year. This position paper updates the roadmap for two of the most promising AI approaches -- reinforcement learning and causal inference - and describes why they could help tip the balance back towards defenders.
\end{abstract}


\section{Introduction}\label{Intro}

\IEEEPARstart{C}{yber} criminality leads to global losses of around \$600 billion per annum \cite{lewis2018economic}. In principle an organisation's security framework is set, implemented and periodically updated by domain experts. Increasing sophistication of cyber-attacks is demanding that these update cycles shorten which creates a need for adaptive and autonomous defences \cite{AntReport}. The potential of artificial intelligence (\ai) to enhance hostile cyber-capability is a further deep concern. An AI controlled attack could be fast, flexible and adaptive and simply overwhelm the current generation of defences. This fact generates an asymmetry between technologies, where defences are usually reactive and lag behind the attacker’s technology \cite{xu2015stochastic}. \ai-based technology is fast becoming a target of cyber-attack in its own right, with many attacks attempting to impact the integrity of the underlying statistical model, in order to make the algorithm misbehave \cite{biggio2018wild}. Such attacks are only going to increase in sophistication. There is a need for a new breed of robust \ai approaches to cyber defence and an increase in the sophistication of active defensive measures. This branch in the cybersecurity literature is called active cyber defence (\acd) \cite{herring2014active,lu2013optimizing,de2019features,xu2015stochastic,ridley2018machine}. The main focus of \acd is to automate defence capability in such a way that it is able to deal with automated threats \cite{herring2014active}. In short, we seek to enhance this automation with \ai approaches.

In this paper we update a recent research roadmap \cite{burke2020} that sets out recommendations for leveraging \ai to develop enhanced \acd capabilities. The roadmap was developed by the Alan Turing Institute, in collaboration with UK's National Cyber Security Centre (\ncsc) and Defence Science and Technology Laboratory (\dstl). In \cref{sec:AI4ACD} we describe the roadmap in more detail and in \cref{RL_section} we describe in more detail a particular research path which involves reinforcement learning (\rl) for \acd{}. We then build on this with causal inference research in \cref{sec:CI_section}, concluding in \cref{sec:conclusion}.

\section{AI for ACD}
\label{sec:AI4ACD}

The main focus of this research initiative is to explore and develop \ai technologies which aim to reduce the asymmetry gap between technologies in which defences just play a reactive role, to provide \ai-based adaptive technologies to power autonomous cyber defence systems \cite{burke2020}. Challenges, which are detailed in \cite{AntReport}, must be overcome in the following three areas:

\begin{itemize}
    \item \ai-enabled network defence -- automatic monitoring, protection and healing of systems;
    \item \ai-enabled security planning -- \ai-enabled planning systems which automatically enhance human decision-making and action taking;
    \item \ai-enabled penetration testing -- \ai agents which automatically identify system weaknesses and vulnerabilities before real-world adversaries.
\end{itemize}

The \acd concept in academic literature presents ambitious requirements for intelligent automation capabilities. In order to create these technologies, researchers and technologists must overcome difficult challenges in \ai, network security and human-machine teaming. Sustained research in fundamental \ai topics applied to realistic cyber scenarios is needed to address the ambitious aims of the \acd concept \cite{burke2020}.  

In the following two sections, we discuss our positional view regarding AI applicability in \acd with two specific projects that were prioritised from those presented in the full roadmap. These projects leverage \rl and causal inference. To aid the exposition of both methodologies, a simulator of an enterprise network was also proposed in \cite{Hoeltgebaum2021}.

\section{Reinforcement Learning}\label{RL_section}

\begin{figure}[!ht]
    \centering
    \includegraphics[width=0.8\columnwidth]{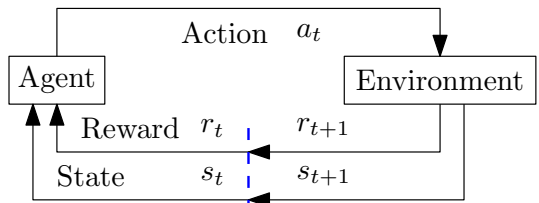}
    \vspace{-1em}
    \caption{Schematic illustration on how \rl works. Figure is an adaptation of the one presented in \cite{sutton1998introduction}.}
    \label{fig:sutton}
\end{figure}

Reinforcement learning (\rl) is based on the idea that an agent learns by interacting with an environment on a trial-and-error  basis. The framework is illustrated in \cref{fig:sutton}, with an agent taking an action and reacting with the environment. Given this action $a_t$, the agent will alter from state $s_t$ to $s_{t+1}$ collecting some reward $r_{t+1}$ (positive or negative) in the process. The process keeps repeating up to a terminal step and is index by time $t$ in the interim. 

The breakthrough in the \rl literature appeared when Google Deepmind made use of \rl agents to beat humans in board and video games \cite{mnih2015human,silver2016mastering,silver2018general}. As a natural consequence, \rl applications in the cyber domain began to emerge. To name a few, recent applications of \rl to set defence strategies, considering an enterprise network, have been explored by \cite{hammar2020finding,elderman2016adversarial,ridley2018machine,bland2020machine,zennaro2020modeling,dowling2018improving}. Note that while \cite{hammar2020finding,elderman2016adversarial,ridley2018machine,bland2020machine,zennaro2020modeling} focused on a round based game, where both attacker and defender play a game in which the attacker tries to conquer a target node and the defender tries to detect it, \cite{dowling2018improving} focused on the use of RL to set honeypots across the network. Building on that, in the sequel we detail our position with experiences acquired when conducting the research in \cite{Hoeltgebaum2021}. Note that an excellent recent survey detailing applications in several cyber security domains making use of deep learning methods with \rl is provided in \cite{nguyen2019deep}.

In \cite{Hoeltgebaum2021} we consider enterprise network security, proposing a highly abstract simulation-based framework to generate scenarios of an enterprise network. The preference for an abstract simulator is motivated by the fact that cyber security data is unlabelled, there is no information about the \emph{ground truth} \cite{heard2018}. In practice this means any information regarding if and when an attack happened is unavailable. Moreover, if real data is used to formulate defence strategies, we would be facing a counterfactual scenario, specifically the question, what would have happened if a specific defence strategy was not applied under specific configurations? Collectively, this leads to the need to simulate data when dealing with cybersecurity applications. This is explicitly stated in \cite{nguyen2019deep} corroborated by \cite{hammar2020finding,zakrzewska2011modeling,elderman2016adversarial,ridley2018machine,bland2020machine,dowling2018improving,bland2020machine,yang2018risk}. These abstractions have the objective of portraying several complexities from a real cyber environment and can provide game like environments where \rl techniques can be deployed to evaluate defence strategies.

The abstraction we proposed in \cite{Hoeltgebaum2021} is a node-based simulation, where each node of the network represents a device with parameters characterising concepts of defence strength (which represents how protected the device is, for instance, security patch updates or anti-virus versions), vulnerabilities and credentials. On the other side of this cyber warfare, the attacker prototype is modeled following lateral traversal attack properties \cite{powell2020epidemiology} with parameters abstracting concepts of attack strength and spread capability. In \cite{Hoeltgebaum2021} the attacker seeks to escalate credentials in an enterprise network. The game ends when the attacker is successful in acquiring what is designated as the \emph{target node}, which may represent, for example the active directory server. Note that with data from the simulator, the defender can understand different properties for controlling defences. Using this abstraction, a stochastic Markov game with one agent is defined in \cite{Hoeltgebaum2021}, representing a centralised defence controller, which acts against an attacker performing lateral traversal. Both attacker and defender have incomplete information and partial observability. The attacker does not know the complete topology of the network. As more nodes are compromised, more information about the target's location is acquired. On the other hand, the defender does know the network topology but not those nodes which are compromised. 
The character of the enterprise cyber-security problem, when expressed as a game, is distinctly different from more familiar game scenarios as we now illustrate with simple examples. 



In the game of chess the rules are fixed throughout the game for both players leading to a comparison of the  players' skills to win the game; in the lateral traversal game, the set of rules is ephemeral and not agreed by both players, specifically the attacker is willing to ignore the rules. This means that the way the game is defined for both players is different (an asymmetric game) in the sense that attacker and defender win in different ways under different set of rules. The attacker seeks to  compromise the network, while the defender can only postpone compromise with a given  set of actions. Additionally, in the cyber environment the set of rules can be adaptive. As a consequence, rewards and optimal policies for both players could potentially be time-varying. After observing a set of movements from the attacker, the defender sets a defence strategy which potentially would not be optimal when facing another kind of attack.

As a second example, consider the famous video game, Pac-Man, which is a maze-based game where an agent must eat the dots inside the maze, while avoid getting caught by ghosts. This game shares the asymmetry of the lateral traversal game in that the players win in different ways. However the rules are fixed for both. Both the agent and the ghosts have a set of rules fixed a priori, movements are limited to left, right, up and down. Given this set of movements, the ultimate goal for the agent is to collect the maximum number of dots to win the game, while not being captured. Making an analogy to the lateral traversal game, we portray the defender as the ghosts trying to catch Pac-Man, who is the attacker. The attacker has a huge range of options and no obligation to follow any rules or policies, or even behave in a consistent manner.


The defender in the lateral traversal game needs to cope with a set of adaptive mechanisms from the attacker while providing best utilisation of enterprise resources - the classic security trade-off between security and usability. Actions the defender takes have costs and consequences. In framing a defence policy, an ideal approach would be that which absorbs these time-varying costs enabling the defender to take decisions in cyber relevant speed while implementing defence strategies. 




One way to reason about decision making challenges for the defender is to propose collaborative defensive agents that are able to exchange information about learned attack patterns. This idea is closely related to the concept of ``benign worms'' (see \cite{lu2013optimizing} and references therein), where these agents continuously traverse the network, interacting with devices, and maintaining the health of the enterprise. These worms would act as defenders taking decisions regarding a computer's safety in real time. However, there are a few drawbacks to discuss, primarily that they create more attack vectors for a malicious entity to exploit. If the worms are controlled by a central agent, the attacker has the opportunity to compromise this agent thereby potentially increasing it’s attack capabilities. A decentralised approach, where each worm is autonomous, reduces the attacker’s potential to compromise them. In either case these worms would require a digital authentication mechanism to validate interacting with a device, opening a new attack vector.


There are significant challenges ahead to construct such defensive agents with learned and adaptive control policies, due to the issues described above, such as asymmetry. A promising direction is provided by a new avenue of research in \rl and control which is already exploring the inclusion of adversarial disturbances in the model \cite{agarwal2019online}. Such approaches could enable defensive agents to collect information from attackers, rather than statistical noise, and then sequentially optimise a sum of revealed cost functions over time. The question we can pose then is, how well can defensive agents perform when compared to rewards achieved under the best stationary policy over time? 


As in the games presented above, real-world attacks are made up of a sequence of malicious actions with several points at which defensive interventions could be made. To address this, we need to supplement our approach with research from the field of causal inference.

\section{Causal Inference}\label{sec:CI_section}



\begin{figure}[ht!]
    \centering
    \includegraphics[width=\columnwidth]{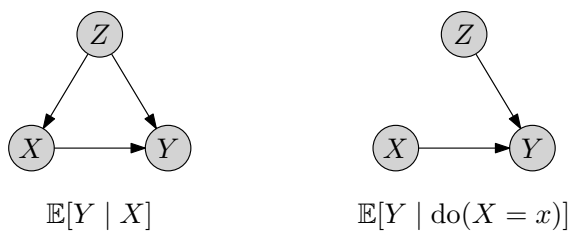}
    \caption{Comparing an observational and interventional distribution, under the intervention $\textrm{do}(X=x)$, where the expected value of the target value ($Y$) is shown w.r.t. the given distribution. In the right panel we are using Pearl's do-operator \cite{pearl2009causality}, graphically applied to the causal graphical model (\cgm).}
    \label{fig:causal_inference}
\end{figure}

In machine learning (\ml) and statistics, usually when we seek to make predictions we do so by estimating conditional distributions, which is to say; we are interested in how variable $Y$ behaves given variable $X$ (and provided we have samples of both). In causal inference we have the same goal, but unlike standard ML we take into account two different kinds of conditional distributions, as indicated in \cref{fig:causal_inference}.

Observational data are samples (measurements) of a system evolving under normal operating conditions. Whereas interventional data are samples of a system evolving under the influence of an intervention acting on the system.  How does causal inference compare to supervised \ml? The observational conditional $p(y \mid x)$ asks ``what is the distribution of $Y$ given that I \underline{observe} that variable $X$ takes value $x$?'' It is a conditional distribution which can simply be calculated as the ratio of two of its marginals. The interventional conditional $p(y \mid \textrm{do}(X=x) )$ asks 
``what is the distribution of $Y$ given that I \underline{set} that variable $X$ takes value $x$?'' This describes the distribution of $Y$ we would observe if we intervened in the data generating process by artificially forcing the variable $X$ to take value $x$, but otherwise simulating the rest of the variables according to the original process that generated the data. This is the causal approach. Observe that the data generating procedure is \emph{not} the same as the joint distribution $p(x,y,z,\ldots)$.

 In the next two sections we will discuss how these simple principles of causality and the do-calculus \cite{pearl2009causality}, can be used for \acd. In particular our focus will be on causal inference applied to the \mitre\footnote{\url{https://attack.mitre.org}} framework. This framework provides definition and classification for a range of cyber-attacks, codified in a consistent and clear manner. It contains a comprehensive matrix of tactics and techniques used by threat hunters and defenders to better classify attacks and assess the risk posed to an organisation from cyber crime. There is however little research into understanding causal links between such tactics and techniques.
 
 The benefits of incorporating the causal aspect are multitude. For example, the causal approach would allow us to understand if the sequence of \mitre tactics are consistent -- not just from a correlation point of view but from a causal standpoint. If a causal structure can be learned from observational and interventional data of a particular attack (say) then, if that pattern is recognised, the defender can amply anticipate the attacker's next move, since we are privy to the causal nature of the attack \cite{qin2004attack}. Indeed the same idea can be entertained for defence strategy.

\subsection{Threat mitigation in the MITRE ATT\&CK framework}

\begin{figure}[ht!]
    \centering
    \includegraphics[width=\columnwidth]{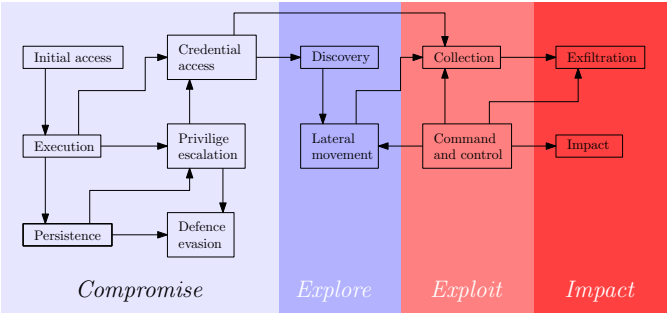}
    \vspace{-2em}
    \caption{Possible causal representations in the \mitre{} framework -- designed by domain experts at the NCSC and DSTL (note: this is not a definite causal representation, there are others). For example, the ``lateral movement'' node corresponds to the adversary trying to move through the environment whereas `collection' means that the adversary is trying to gather data of interest to their goal. The diagram is encoded with a colour gradient which from left to right, blue to red, signifies the potential damage an intruder will cause by engaging in those specific techniques (nodes).}
    \label{fig:mitre}
\end{figure}

The framework is used as a foundation for the development of specific threat models and methodologies in the private sector, in government, and in the cyber security product and service community. Therein there are multiple headings under which techniques are listed according to what an adversary is trying to achieve, see \cref{fig:mitre}.

\begin{figure}[hb!]
    \centering
    \includegraphics[width=\columnwidth]{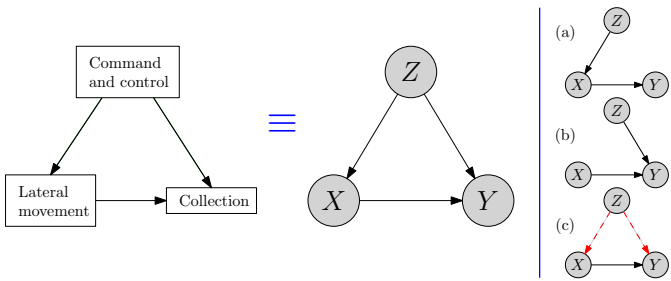}
    \vspace{-2em}
    \caption{Sub-graph (selection of tactics from \cref{fig:mitre}) of MITRE framework with variable labelling for tactics used in this toy example. The shaded node representation means that this tactic is ``observed'' or ``can be measured''. The right column denotes possible derivative topologies.}
    \label{fig:mitre_small}
\end{figure}

 We seek to analyse the larger \mitre framework through the lens of a small subset of the available techniques, found in \cref{fig:mitre}, investigating, in depth, their causal relationships.  Specifically, we seek to understand the causal relationships across time. The toy-example (a small sub-\cgm of the full \cgm found in \cref{fig:mitre}) is shown in \cref{fig:mitre_small}.

There are multiple ways in which the topologies can unfold over time, and some of these are represented in \cref{fig:mitre_small} -- in (b) the manipulative variables (interventions) $Z$ and $X$ are, in the spatial setting, fully independent. One intervention does not affect the other in that time-slice. In (a) there is a dependence between interventions; $Z$ naturally has to flow through $X$ in order to affect the target variable $Y$. But further, in (c) the connections change over time to simulate a scenario where some external factor(s) are influencing the unfolding of the dynamics in the system.

Now, consider the application side of the three scenarios. In scenario (a) lateral movement is a function of a command and control attack. Once the malware has gained entry to the system, the attack will typically evolve through the different stages of the kill chain. It carries out early reconnaissance, creates a state of persistence, seeks access to the outside world through a C\&C server and then initiates a series of lateral movements, until it reaches its final goal of data exfiltration (other impacts are possible as well of course). In (b) the intruder is telling the malware what to collect, but the malware moves around the network without direct control. Finally in (c) the CGM has a confounder node that represents an attacker who can directly control both the lateral movement and/or the collection. Which is to say that an adversary may search local system sources, such as file systems or local databases, to find files of interest and sensitive data prior to exfiltration.


There are of course multiple scenarios we could entertain but for the sake of brevity we settle for these three. The graph structures that we investigate can be motivated by some known malware structure found in \cite[fig. 7]{lee2014gmad}. Finally, the time-index version of these manipulative and target variables are connected in a dynamical Bayesian network, and this particular set of nodes are chosen because they straddle the exploration/exploitation boundary. The analysis thus far hinges on us being privy to a present threat to the system. Which is to say that this type of threat mitigation is not useful without threat detection.

\subsection{Threat detection}

\begin{figure}[ht!]
    \centering
    \includegraphics[width=\columnwidth]{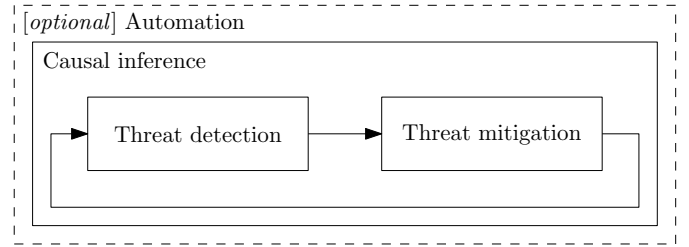}
    \vspace{-2em}
    \caption{Causal inference learning scheme in a cyber security context. Malware or threat detection (left box) box precedes mitigation of the threat (right box), with the found intervention being returned to the threat detection in the form of feedback, to enable better detection in the future. In this abstraction and interpretation, the scheme sits within an optional automation framework, meaning that the user selects the desired level of autonomy.}
    \label{fig:threat_detection}
    \vspace{-1em}
\end{figure}

In an offline setting where we are privy to all the observational data and have a clear understanding of causal associations in our system (i.e. we have a good approximation or indeed the true, causal graphical model) we seek to understand if a suite of \mitre indicators of compromise (e.g. suspicious file hashes or IP addresses) constitute malicious or benign activity. From a signal processing stance this would constitute a smoothing problem in which we can perform inference from the back and front of the graph. Thus, given a set of temporal features (where we assume that indicators of compromise are temporal and have been logged across time), can we classify this sequence as benign or malign? This falls under the topic of time-series classification. A successful threat detection would then inform the threat mitigation approach as shown in \cref{fig:threat_detection}, and any interventions would then change the next set of observations.

To enable true breakthroughs in threat detection, in addition to the proposed simulator, the research community needs enhanced approaches to simulating data to provide realistic scenarios for meaningful threat detection, ideally based on real observations from enterprises or from complex cyber arenas.

\section{Final remarks}
\label{sec:conclusion}

We have discussed some interesting challenges and prospects for the development of AI for ACD in the light of the complexities of enterprise cyber defence. Promising work from the literature has been discussed and we suggest directions for further development. 

A potentially rich avenue of further research arises by combining \cref{RL_section} and \cref{sec:CI_section} performing \emph{causal reinforcement learning}. Such approaches have recently gained traction in the research community \cite{BareinboimCRL,zhang2020designing} and can provide several insights into  how to set controlling defence strategies for specific types of attacks. We also refer to their online repository for tutorials and references \cite{CRLwebpage}. 


Making substantial progress on our roadmap will require sustained multi-year investment from government and industry, as well a range of new collaborative research initiatives focussed on developing game-changing AI-based defence capabilities.





\vspace{-0.2cm}
\bibliographystyle{IEEEtran}
\bibliography{Bibliography}




\end{document}